\documentclass[useAMS,usenatbib]{mn2e}
\usepackage{graphicx}
\usepackage[dvips]{color}
\usepackage{txfonts}
\usepackage{multirow}
\usepackage{amsfonts}
\usepackage{amssymb}
\usepackage{natbib}

\newcommand{\te}{$T_{\rm eff}$}
\newcommand{\logg}{$\log{g}$}
\newcommand{\vsini}{$v\sin{i}$}
\newcommand{\vmicro}{$v_{\rm turb}$}
\newcommand{\kms}{km\,s$^{-1}$}
\newcommand{\cd}{d$^{-1}$}
\newcommand{\mhz}{$\mu$Hz}

\title[Modelling of the $\sigma$~Sco binary system]{Modelling of $\sigma$\,Scorpii, a high-mass binary with a $\beta$\,Cep variable primary component}

\author[A. Tkachenko et al.]{A.\ Tkachenko,$^{1,}$\thanks{Postdoctoral Fellow of the Fund for Scientific Research (FWO), Flanders, Belgium} C.\ Aerts,$^{1,2}$ K.\ Pavlovski,$^3$ P.\ Degroote,$^{1,}$\thanks{Postdoctoral Fellow of the Fund for Scientific Research (FWO), Flanders, Belgium} P.~\ I.\ P\'{a}pics,$^1$ \newauthor
        E.\ Moravveji,$^1$ H.\ Lehmann,$^4$ V.\ Kolbas,$^3$ and K.\ Cl\'{e}mer$^1$\\
  $^1$Instituut voor Sterrenkunde, KU Leuven, Celestijnenlaan 200D, B-3001 Leuven, Belgium\\
  $^2$Department of Astrophysics, IMAPP, Radboud University Nijmegen, 6500 GL Nijmegen, The Netherlands\\
  $^3$Department of Physics, University of Zagreb, Bijeni\v{c}ka cesta 32, 10000 Zagreb, Croatia\\
  $^4$Th\"{u}ringer Landessternwarte Tautenburg, 07778 Tautenburg, Germany}

\date{Received date; accepted date}

\pagerange{\pageref{firstpage}--\pageref{lastpage}} \pubyear{2014}

\begin{document}

\label{firstpage}

\maketitle

\begin{abstract}
  High-mass binary stars are known to show an unexplained discrepancy between
  the dynamical masses of the individual components and those predicted by models. In this work, we study
  Sigma Scorpii, a double-lined spectroscopic binary system
  consisting of two B-type stars residing in an eccentric orbit. The
  more massive primary component is a $\beta$~Cep-type pulsating variable star. Our analysis is based on a time-series of some 1\,000 high-resolution spectra collected with the {\sc coralie}
  spectrograph in 2006, 2007, and 2008. We use two different approaches to
  determine the orbital parameters of the star; the spectral
  disentangling technique is used to separate the spectral
  contributions of the individual components in the composite spectra. The non-LTE based
  spectrum analysis of the disentangled spectra reveals two stars of
  similar spectral type and atmospheric chemical composition.
  Combined with the orbital inclination angle estimate found in the
  literature, our orbital elements allow a mass estimate of
  14.7$\pm$4.5~M$_{\odot}$ and 9.5$\pm$2.9~M$_{\odot}$ for the
  primary and secondary component, respectively. The primary component is found to pulsate in three independent
  modes, of which two are identified as fundamental and second overtone radial modes, while the third is an $l=1$ non-radial mode. Seismic modelling of the pulsating component refines stellar parameters to 13.5$^{+0.5}_{-1.4}$~M$_{\odot}$ and 8.7$^{+0.6}_{-1.2}$~M$_{\odot}$, and delivers radii of 8.95$^{+0.43}_{-0.66}$~R$_{\odot}$ and 3.90$^{+0.58}_{-0.36}$~R$_{\odot}$ for the primary and secondary, respectively. The age of the system is estimated to be
  $\sim$12~Myr.
\end{abstract}


\section{Introduction}

Sigma Scorpii ($\sigma$~Sco, HD\,147165, HR\,6084) is a quadruple
system consisting of a close double-lined spectroscopic binary, and
two other fainter stars at distances of about 0.4$^{\prime\prime}$
and 20$^{\prime\prime}$ \citep[][and references therein]{North2007}.
The close pair consists of a B1\,III $\beta$~Cep-type variable star
(hereafter called primary) and a B1\,V companion (hereafter called
secondary). The two stars reside in an eccentric orbit; the orbital
period of the system is $\sim$33~days.

After the discovery of the radial velocity (RV) variations in the
system by \citet{Slipher1904}, the star was subjected of numerous
spectroscopic, photometric and interferometric studies. Two
spectroscopic studies by \citet{Levee1952} and \citet{Struve1955}
gave consistent results in the sense that both found variability
intrinsic to the B1 giant star with the dominant period being about
0.25~days. Both studies reported about variability in the
$\gamma$-velocity which was interpreted as evidence of a companion
star. The orbital solution obtained by \citet{Levee1952} and
\citet{Struve1955} independently revealed an orbital period of about
33~days, though very different eccentricities of 0.2 and 0.36,
respectively. In both cases, the dominant period of $\sim$0.25~d was
attributed to stellar pulsations. \citet{Struve1955} additionally
reported a period increase with the rate of 2.3~s\,cen$^{-1}$. Using
the RVs collected by both authors, \citet{Fitch1967} classified the
star as a single-lined spectroscopic binary and reported an orbital
solution in agreement with the findings by \citet{Struve1955},
including a high eccentricity value of $\sim$0.4. \citet{Osaki1971}
argued, however, that the variability of the $\gamma$-velocity
reported by \citet{Levee1952} and \citet{Struve1955} is not
necessarily connected to the existence of a stellar companion but
might be related to non-radial pulsations.

Stars are expected to show slow changes in their dominant pulsation
period in the course of the hydrogen exhaustion phase, but it is a
matter of debate whether such an evolutionary effect is detectable.
After the first report on the period change of the fundamental mode
of the primary of $\sigma$\,Sco by \citet{Struve1955}, a subsequent
period decrease and increase was detected by \citet{Hoof1966} and
\citet{Sterken1975}, respectively. The latter study also reported on
a periodicity of the observed variations with a period close to 23
years.

\citet{Kubiak1980} presented one of the first detailed periodicity
studies for $\sigma$\,Sco based on the RVs obtained by
\citet{Henroteau1925}, \citet{Levee1952}, and \citet{Struve1955}. In
total, a time-series of 862 measurements has been analysed and
revealed eight frequency peaks ranging in amplitude from 1 to 40
(for the dominant mode) \kms. Two out of the eight frequencies, the
dominant mode f$_1$=4.05111~\cd\ (46.87132~\mhz) and
f$_2$=4.17588~\cd\ (48.31495~\mhz), were found to be independent
modes whereas the other six frequencies appeared as their low-order
combination terms (including harmonics). Based on the analysis of
both photoelectric (taken in $ubvy$ Stromgren system) and
spectroscopic data, \citet{Kubiak1983} found non-radial pulsations
to be the most reasonable explanation of the observed variability.

An extensive photometric study of the system was presented by
\citet{Goossens1984} based on the observations collected by
\citet{Hoof1966} in the ultraviolet (UV), yellow (Y), and blue (B)
light. The authors reported on the light variability in all three
pass-bands with the same two dominant modes as found by
\citet{Kubiak1983}. On the other hand, the colour variations
revealed only one dominant frequency f$_1$=4.05118~\cd\
(46.87215~\mhz). The analysis performed by \citet{Goossens1984} on
the RVs collected by \citet{Levee1952} and \citet{Struve1955}
appeared to be consistent with the analysis of the colour
variations. The authors also found the dominant mode to be variable
both in amplitude and phase with a period of $\sim$8.3~days, i.e., a
quarter of the orbital period. Two models capable of representing
the light variability intrinsic to the primary component were
proposed: i) two intrinsic pulsation modes with constant periods and
amplitudes (in agreement with \citealt{Struve1955} and
\citealt{Hoof1966}); and ii) one intrinsic oscillation of which both
the amplitude and the mean light curve are modulated by tidal action
(supported by the results of the analysis of colour and RV
variations).

\begin{table}
\tabcolsep 1.5mm\caption{Orbital solution for $\sigma$\,Sco as
derived by \citet[][indicated as M1991]{Mathias1991},
\citet[][indicated as P1992]{Pigulski1992}, and \citet[][indicated
as N2007]{North2007}. Errors are given in parentheses in terms of
last digits. Subscripts 1 and 2 refer to the primary and secondary
component, respectively.}\label{Table1}
\begin{tabular}{lllll} \hline
Parameter & Unit & M1991 & P1992 & N2007\\\hline
$P$ & days & 33.012(2) & 33.012 & 33.010(2)\\
$K_1$ & \kms & 31.0(1.3) & 29.2(1.5) &\\
$K_2$ & \kms & 40.3(7.2) & &\\
$e$ & & 0.44(11) & 0.40(4) & 0.3220(12)\\
$\omega$ & degrees, $^{\circ}$ & 299.0(10.0) & 287.0(6.0) & 283.0(5.0)\\
$\gamma$ & \kms & 1.88(1.25) & 3.9(8) &\\
T$_0$ & HJD (2\,434\,000+) & 889.4(7) & 888.0(4) & 889.0(1.0)\\
a$_1\sin i$ & (10$^6$) km & 13.0(2.4) & 12.2(6) &\\
a$_2\sin i$ & (10$^6$) km & 20.2(4.1) & &\\
$f$(M$_1$) & M$_{\odot}$ & 0.08(5) & 0.066(10) &\\
$f$(M$_2$) & M$_{\odot}$ & 0.30(18) & &\\\hline
\end{tabular}
\end{table}

\citet{Mathias1991} investigated the $\sigma$\,Sco system based on
newly obtained time-series of high-resolution, high signal-to-noise
ratio (S/N) \'{e}chelle spectra. The authors reported the detection
of the lines of the companion star in their spectra, and presented
the orbital solution as summarised in Table~\ref{Table1}. The Van
Hoof effect that stands for a small phase lag of about 0.04$P$ of
the RV curve of the hydrogen lines behind all other spectral lines,
was detected in the system for the first time. \citet{Pigulski1992}
presented another study of the $\sigma$\,Sco system based on all
photometric and spectroscopic data available at that time. The
author determined the orbital parameters of the system as listed in
Table~\ref{Table1} and investigated the $O-C$ diagram of the main
pulsation mode. The latter analysis revealed variability in the
period of the dominant mode which \citet{Pigulski1992} explained in
terms of a superposition of the light-time effect due to a third
body in the system and an evolutionary increase of the intrinsic
pulsation period with a rate of 3~s\,cen$^{-1}$. These results are
consistent with the early findings by \citet{Struve1955}. Using the
same data set as \citet{Pigulski1992} did, \citet{Cugier1992}
performed a mode identification of the two dominant pulsation modes
f$_1$=4.05121~\cd\ (46.87250~\mhz) and f$_2$=4.17240~\cd\
(48.27467~\mhz). The authors found that f$_1$ is compatible with a
radial mode whereas f$_2$ is most probably an $l$=2 mode. The
identification for f$_1$ was confirmed by \citet{Heynderickx1994}
later on, based on the information on the wavelength dependence of
the photometric amplitudes.

\begin{table}
\centering \tabcolsep 2.5mm\caption{List of the spectroscopic
observations of $\sigma$\,Sco. JD is the Julian Date, $N$ gives the
number of spectra taken during the corresponding observational
period.}\label{Table2}
\begin{tabular}{llr} \hline
\multicolumn{2}{c}{Time period} & $N$\\
\multicolumn{1}{c}{Calendar date} & \multicolumn{1}{c}{JD
(2\,450\,000+)} &\\\hline
11.03--11.03.2006 & 3805.82--3805.87 & 7\\
14.03--18.03.2006 & 3808.78--3812.86 & 46\\
05.04--05.04.2006 & 3830.79--3830.91 & 17\\
08.04--09.04.2006 & 3833.77--3834.92 & 26\\
12.04--12.04.2006 & 3837.70--3837.85 & 19\\
31.05--31.05.2006 & 3886.59--3886.86 & 6\\
04.06--06.06.2006 & 3890.59--3892.62 & 6\\
08.06--08.06.2006 & 3894.56--3894.83 & 21\\
10.06--17.06.2006 & 3896.70--3903.83 & 134\\
27.07--30.07.2006 & 3944.49--3946.72 & 14\\
05.08--15.08.2006 & 3953.49--3962.62 & 326\\
19.03--23.03.2007 & 4178.90--4182.92 & 22\\
25.03--26.03.2007 & 4184.90--4185.92 & 8\\
30.03--31.03.2007 & 4189.91--4190.93 & 9\\
18.04--22.04.2007 & 4208.87--4212.95 & 92\\
25.04--26.04.2007 & 4215.88--4216.95 & 26\\
28.04--29.04.2007 & 4218.85--4219.94 & 34\\
22.07--28.07.2007 & 4303.57--4309.70 & 136\\
30.07--31.07.2007 & 4311.59--4312.59 & 41\\
13.06--15.06.2008 & 4630.54--4632.78 & 30\\\cline{3-3}
\multicolumn{2}{l}{Total number of spectra:\rule{0pt}{11pt}} & {\bf
1020}\\ \hline
\end{tabular}
\end{table}

\citet{North2007} derived the orbital solution and physical
parameters of the $\sigma$\,Sco system based on interferometric
data. The authors found a significantly lower eccentricity of about
0.32, whereas the other three orbital parameters (period $P$, time
of periastron passage T$_0$, and longitude of periastron $\omega$)
agreed within the errors with those reported by \citet{Mathias1991}
and \citet{Pigulski1992} (cf. Table~\ref{Table1}). Using the RV
semi-amplitudes, K$_{1}$ and K$_{2}$, determined by
\citet{Mathias1991}, \citet{North2007} estimated the masses to be
M$_1$=18.4$\pm$5.4 and M$_2$=11.9$\pm$3.1~M$_{\odot}$, and assigned
spectral types B1\,III and B1\,V to the primary and secondary
components, respectively. Following the spectral type and luminosity
class, an estimate of the effective temperature of the secondary of
25\,400$\pm$2\,000~K was presented for the first time.

Despite these efforts made by different research groups, no precise
fundamental parameters for both binary components are available. For
example, the effective temperature estimates for the $\beta$\,Cep
pulsating primary component range from 21\,880 \citep{Beeckmans1977}
to 30\,000~K \citep{Theodossiou1985} (see Table~4 in
\citealt{Vander_Linden1988} for an overview of the primary
fundamental parameters estimates to which the values of
\te=25\,700$\pm$1\,500~K and \logg=3.85 derived by
\citealt{Niemczura2005} have to be added). On the other hand, only
one rough estimate of the temperature of the secondary by
\citet{North2007} is available in the literature. The \logg\ of the
primary is not well constrained either, ranging from 3.5
\citep{Underhill1979} to 4.0~dex \citep{Schild1971,Heasley1982}. No
estimate of this parameter exists for the secondary. Moreover, a
possible influence of the large-amplitude ($\sim$80~\kms,
\citealt{Kubiak1980}) radial pulsation mode on the orbital solution
has been ignored in all previous studies, which might have a
significant impact on the determined orbital and physical parameters
of the system. This motivated us to start this work, with the aim to
determine precise orbital and stellar parameters of the system,
evaluate atmospheric chemical composition and deduce the current
evolutionary stage of both binary components, ideally from
asteroseismic modelling.

\begin{figure}
\includegraphics[scale=0.85]{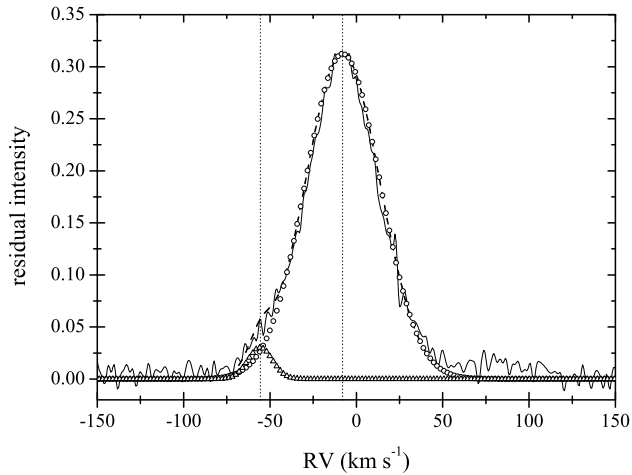}
\caption{The composite profile of Si\,{\small III}\,4552~\AA\
spectral line observed at JD~2454311.6368 (solid line) with the best
fit double-Gaussian profile overplotted (dashed line). The vertical
dotted lines mark the derived RVs for both components. The
individual Gaussians are shown by open circles and triangles for the
primary and secondary, respectively.} \label{Figure1}
\end{figure}

We present our newly obtained high-quality spectroscopic data as
well as describe the basic steps of the data reduction in Sect.~2.
Sect.~3 is devoted to the orbital solution which we obtained by
means of an iterative method. The orbital elements and the analysis
of separated spectra obtained with the spectral disentangling
technique are presented in Sect.~4. System and physical parameters
of $\sigma$~Sco are discussed in Sect.~5. We present the results of
the spectroscopic mode identification in Sect.~6. Stellar modelling
is described in Sect.~7. The paper closes with a discussion and
conclusions presented in Sect.~8.

\section{Observations and data reduction}

We obtained a time-series of 1020 high-resolution, high S/N spectra
with the {\sc coralie} spectrograph attached to the 1.2m Euler Swiss
Telescope (La Silla, Chile). The data were gathered during three
consecutive years, in 2006, 2007, and 2008. The spectra have a
resolving power of R=50\,000 and cover a 3000~\AA\ wide wavelength
range, from 381 to 681~nm. Table~\ref{Table2} gives the log of our
spectroscopic observations and lists the period of observations
(both calendar and Julian dates) and the number of acquired spectra
during the corresponding period.

The data reduction has been completed using a dedicated pipeline and
included bias and stray-light subtraction, cosmic rays filtering,
flat fielding, wavelength calibration by ThAr lamp, and order
merging. The continuum normalization was done afterwards by fitting
a (cubic) spline function through some tens of carefully selected
continuum points. More information on the normalization procedure
can be found in \citet{Papics2012}.

As it is mentioned in the introduction, $\sigma$~Sco is in fact a
quadruple system with one of the components being a triple star.
This three stellar component system consists of a double-lined
spectroscopic binary (the one we study in this paper) and a 2.2~mag
fainter tertiary at a distance of about 0.4$^{\prime\prime}$ from
the close pair. Despite this proximity of the third component, no
signature of it could be detected in our composite spectra.
Moreover, a compilation of the speckle interferometric observations
of $\sigma$~Sco presented by \citet{Pigulski1992} shows that only a
small part of the tertiary orbit was covered within $\sim$12~yr.
\citet{Nather1974} and \citet{Evans1986} suggested the orbital
period of the order of 100-350~yr from the analysis of the lunar
occultations of $\sigma$~Sco. This means that the orbital period of
the tertiary is much larger than the time base of our observations.
As such, the third component does not affect our spectroscopic
analysis of the close pair, and we neglect it in our study.

\section{Orbital solution: iterative method}

\begin{figure}
\includegraphics[scale=0.75]{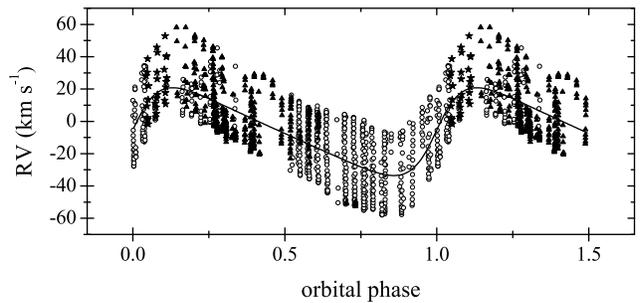}
\caption{RVs of the pulsating primary of the $\sigma$\,Sco system
(symbols) folded with the orbital period of 33.076~d (see text).
Open circles, filled triangles, and stars refer to the data from
2006, 2007, and 2008, respectively. The best fit is shown with the
solid line.} \label{Figure2}
\end{figure}

In this section, we describe the way RVs of both components of
$\sigma$\,Sco were determined from the composite spectra and present
our first attempt to compute precise orbital solution by means of an
iterative procedure.

\subsection{Calculation of the RVs}

The RVs of both binary components were computed based on the
observed composite profiles of the Si\,{\small III}\,4552~\AA\
spectral line. The observations were fitted with a superposition of
two Gaussian profiles; the free parameters were RV, full width at
half maximum (FWHM), and the strength for each of the two profiles.
Fig.~\ref{Figure1} illustrates an example of the double-Gaussian fit
to one of the composite line profiles of the $\sigma$~Sco binary.
The derived RVs of both components are indicated by the vertical
dotted lines; the fit itself is shown by the dashed line.

Determination of the RVs of both stellar components was possible
only for approximately a quarter of all the spectra we had at our
disposal. In all other cases, the RVs of the secondary component
could not be measured, lacking the detection the companion's
contribution in the observed composite spectra of the binary. The
detection of the lines of the secondary, in most cases, was
prevented by the large-amplitude spectral variability due to the
radial pulsation mode of the primary component. This variability
manifests itself in terms of large periodic shifts of the lines of
the primary in velocity space, which means that the spectra taken at
the same orbital phase look very different depending on the phase at
which the pulsation mode of the primary is caught. After a careful
inspection, we found that even those RVs of the secondary which we
could measure from the composite profiles were significantly
affected by the radial mode of the primary. Thus, we computed our
orbital solution in a single-lined binary mode, relying on the RVs
of the primary only. The RVs are shown in Fig.~\ref{Figure2} along
with the obtained orbital solution (see below).

\begin{table}
\tabcolsep 3.0mm\caption{Orbital elements computed based on the RVs
of the primary component. The two solutions are based on the
original and the prewhitened data sets (see text for details).
Errors are given in parentheses in terms of last
digits.\label{Table3}}
\begin{tabular}{llll}
\hline
& & \multicolumn{2}{c}{Data set}\\
Parameter & Unit & \multicolumn{1}{c}{Original} & \multicolumn{1}{c}{Prewhitened}\\
\hline
$P$ & d & 33.076(64) & 33.016(12)\\
$K_1$ & \kms & 26.89(1.08) & 30.14(35)\\
$e$ & & 0.359(187) & 0.333(61)\\
$\omega$ & degrees, $^{\circ}$ & 272.7(6.6) & 281.4(1.6)\\
$T_0$ & HJD (2\,434\,000+) & 881.4(3.8) & 885.8(1.0)\\
$\gamma$ & \kms & --8.80(86) & -1.52(25)\\\hline
{\sc rms} & \kms & 16.7614 & 1.8556\\
\hline
\end{tabular}
\end{table}

\subsection{Orbital elements}

For the calculation of the orbit of the primary component, we used a
computer program that is based on the method of differential
corrections to the orbital elements \citep[see,
e.g.,][]{Schlesinger1910,Sterne1941} and was written by one of us
(HL). We optimized for the following six orbital elements
simultaneously: the orbital period $P$, the time and the longitude
of the periastron ($T_0$ and $\omega$, respectively), the
semi-amplitude ($K_1$) of the primary and gamma-velocity ($\gamma$)
of the system, and the orbital eccentricity $e$. Fig.~\ref{Figure2}
illustrates the initial fit to the original data, where a large
``RV-scatter'' is noticeable. This short-term variability is clearly
intrinsic to the primary and is consistent with the radial pulsation
mode with a frequency of $\sim$4.05~\cd\ (46.86~\mhz) reported in
the previous studies. The derived orbital elements are listed in
Table~\ref{Table3} (third column, indicated as ``Original''). The
elements differ significantly from those reported in
\citet{Mathias1991} and \citet{North2007} who, in particular, found
a significantly shorter orbital period and higher semi-amplitude of
the primary, as well as significantly different $\gamma$-velocity
and eccentricity of the system.

Apparently, one needs to get rid of the large amplitude intrinsic
variability of the primary component to obtain a reliable orbital
solution. One way to minimize the influence of the pulsations on the
orbital solution is to introduce an iterative procedure which
comprises of three basic steps: i) calculation of the orbital
solution and subtracting it from the original data set; ii)
searching for the periodicities in the obtained residuals; and iii)
subtracting the pulsation signal from the original data set and
re-calculating the orbital solution. Our further analysis in this
section is based on such a procedure which is repeated until no
significant changes in the derived orbital elements and extracted
frequencies occur anymore.

\begin{table}
\tabcolsep 2.3mm\caption{\small Frequencies found in the combined
(residual) data set from 2006, 2007, and 2008. The frequency
uncertainties are given by the Rayleigh limit which amounts to
0.0019~\cd\ (0.0219~\mhz). The amplitude errors are given in
parentheses in terms of last digits.}
\begin{tabular}{lr@{.}lr@{.}lr@{.}lr@{.}l}
\hline
f$_{\rm i}$ & \multicolumn{4}{c}{Frequency} & \multicolumn{2}{c}{Amplitude} & \multicolumn{2}{c}{S/N}\\
 & \multicolumn{2}{c}{\cd} & \multicolumn{2}{c}{\mhz} & \multicolumn{2}{c}{\kms} & \\
\hline
f$_1$   & 4&0515 & 46&8759 & 23&6(1) & 144&8\\
f$_2$=2f$_1$   & 8&1030 & 93&7517 & 2&0(5) & 12&4\\
f$_3$   & 4&1726 & 48&2770 & 1&4(1) & 9&2\\
f$_4$=3f$_1$   & 12&1521 & 140&5998 & 1&1(2) & 7&9\\
f$_5$   & 5&9706 & 69&0201 & 1&0(2) & 5&5\\
f$_6$   & 0&5003 & 5&7885 & 0&7(2) & 4&1\\
\hline
\end{tabular}
\label{Table4}
\end{table}

\begin{table}
\tabcolsep 2.3mm\caption{\small Evolution of the amplitudes of the
two dominant modes and harmonics of the radial mode through years
2006--2008. The amplitude errors are given in parentheses in terms
of last digits.}
\begin{tabular}{lr@{.}lr@{.}lr@{.}l}
\hline
f$_{\rm i}$ & \multicolumn{6}{c}{Amplitude (\kms)}\\
 & \multicolumn{2}{c}{2006} & \multicolumn{2}{c}{2007} &
 \multicolumn{2}{c}{2008}\\
\hline
f$_1$   & 23&6(12) & 22&1(3) & 20&7(4)\\
f$_2$=2f$_1$ & 2&1(6) & 2&0(6) & 1&7(7)\\
f$_3$   & 1&7(3) & 1&5(3) & 1&4(4)\\
f$_4$=3f$_1$ & 1&3(3) & 1&1(4) & 0&9(5)\\ \hline
\end{tabular}
\label{Table4a}
\end{table}

\begin{figure}
\includegraphics[scale=0.75]{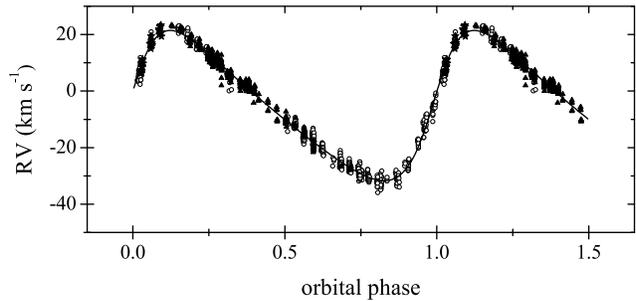}
\caption{RVs of the primary of $\sigma$\,Sco after prewhitening from
six frequencies listed in Table~\ref{Table4}. Open circles, filled
triangles, and stars refer to the data from 2006, 2007, and 2008,
respectively. The solid line shows the final orbital solution; the
corresponding orbital elements are given in Table~\ref{Table3}
(fourth column, indicated as ``Prewhitened'').} \label{Figure3}
\end{figure}

We used the {\sc Period04} program \citep{Lenz2005} to search for
the periodocities in the residuals. The decision on the significance
of the extracted frequencies was made according to the significance
level of S/N=4.0 proposed by \citet{Breger1993} for ground-based
data. The noise level was computed from a 3~\cd\ box after
prewhitening the frequency peak in question. Table~\ref{Table4}
lists all frequencies detected in our data assuming the stop
criterion mentioned above. To verify our solution at each
prewhitening step, the corresponding residuals were compared to the
previous data set from which the frequency in question has been
subtracted. We found that the first four frequencies listed in
Table~\ref{Table4} comprise the major part of the short-term
variability present in the RVs of the primary of $\sigma$~Sco. A
further prewhitening of the data from frequencies f$_5$ and f$_6$
leads to only a marginal improvement of the solution, by
$\sim$0.15~\kms\ in terms of the {\sc rms}. Our final orbital
solution obtained after prewhitening the six frequencies listed in
Table~\ref{Table4} is given in the last column of
Table~\ref{Table3}; the quality of the fit is illustrated in
Fig.~\ref{Figure3}. Frequencies f$_1$, f$_3$, and f$_5$ have already
been reported in several previous studies \citep[see,
e.g.,][]{Kubiak1980,Jerzykiewicz1984,Cugier1992}, two frequencies,
f$_2$ and f$_4$, appear to be the second and the third harmonics of
the dominant mode, respectively, and frequency f$_6$ is a new
detection pending an independent confirmation. Using the inclination
angle reported by \citet{North2007} and our estimates of the radius
of the primary (cf. Sect.~5), we find that f$_6$ is by
$\sim$0.3~\cd\ larger than the expected rotational frequency of this
star.

\begin{table}
\tabcolsep 2.7mm\caption{Orbital solutions obtained by means of the
spectral disentangling technique based on two different sets of
spectra (see text for details). Errors of measurement are 1$\sigma$
standard deviations and are given in parentheses in terms of last
digits.\label{Table5}}
\begin{tabular}{llll}
\hline
& & \multicolumn{2}{c}{Data set}\\
Parameter & Unit & \multicolumn{1}{c}{Orbit} & \multicolumn{1}{c}{Orbit+Spectra}\\
\hline \multicolumn{4}{c}{{\bf Solution 1}}\\
$P$ & d & 33.016(12)$^*$ & 33.016(12)$^*$\\
$K_1$ & \kms & 24.32(12) & 25.31(13)\\
$K_2$ & \kms & 51.03(93) & 49.70(50)\\
$e$ & & 0.334(2) & 0.334(2)$^{**}$\\
$\omega$ & degrees, $^{\circ}$ & 289.0(8) & 282.0(5)\\
$T_0$ & HJD (2\,434\,000+) & 885.17(6) & 884.63(5)\vspace{5mm}\\
\multicolumn{4}{c}{{\bf Solution 2}}\\
$P$ & d & 33.016(12)$^*$ & 33.016(12)$^*$\\
$K_1$ & \kms & 30.14(35)$^*$ & 30.14(35)$^*$\\
$K_2$ & \kms & 41.86(1.25) & 47.01(98)\\
$e$ & & 0.383(8) & 0.383(8)$^{**}$\\
$\omega$ & degrees, $^{\circ}$ & 294.4(1.7) & 288.1(8)\\
$T_0$ & HJD (2\,434\,000+) & 886.42(12) & 886.11(4)\\\hline
\multicolumn{4}{l}{$^*$ fixed to the value obtained in Sect.~3}\\
\multicolumn{4}{l}{$^{**}$ fixed to the value from solution
``Orbit''}
\end{tabular}
\end{table}

In order to check for the stability of the frequencies and their
amplitudes in time, we analysed the data sets from individual years
separately. Given that the data subsets from 2007 and 2008 are
shorter and contain less measurements than the one from 2006, it is
not surprising that only the dominant mode at $\sim$4.05~\cd\
(46.86~\mhz) and its harmonics, and the mode at f$_3$ $\simeq$
4.17~\cd\ (48.25~\mhz) could be detected in the corresponding data
subsets. Our analysis revealed a small decrease of about
6$\times$10$^{-5}$~\cd\ (7$\times$10$^{-4}$~\mhz) of the frequency
of the dominant mode between the years 2006 and 2007, which is two
orders of magnitude smaller than the actual frequency resolution.
The frequency resolution for the data taken in 2008 is too low to
draw firm conclusions on the period increase/decrease during this
season. A small but significant amplitude decrease of the two
dominant modes and harmonics of the radial mode was also detected
through the years 2006--2008 (see Table~\ref{Table4a}). This is
consistent with the recent study by \citet{Handler2014} who also
reported a pulsation amplitude decrease in the primary of
$\sigma$\,Sco, but at a larger rate. The short time base of our
observations prevents us from drawing any firm conclusions with
respect to the physical cause of the detected amplitude decrease.
Finally, we note that the frequency of the dominant mode of
4.0515~\cd\ (46.8759~\mhz) detected by us from the combined data set
is different from the one of 4.0512~\cd\ (46.8724~\mhz) reported by
\citet{Pigulski1992}. This difference is still an order of magnitude
smaller than provided by the frequency resolution of our data to
conclude on any period changes of the dominant oscillation mode.

\section{Spectral disentangling}

The orbital solution obtained in Sect.~3 is not perfect but provides
a good starting point for the analysis with more sophisticated
methods like the spectral disentangling ({\sc spd}) technique. In
the method of {\sc spd}, as introduced by \citet{Simon1994}, one
simultaneously solves for the individual spectra of stellar
components of a multiple system and a set of the orbital elements.
\citet{Hadrava1995} suggested an implementation of the technique in
Fourier space, which significantly decreases the computation time.
In this work, we apply the {\sc spd} technique in Fourier space as
implemented in the {\sc FDBinary} code \citep{Ilijic2004}.

\subsection{Orbital solution}

All existing Fourier-based {\sc spd} codes (including {\sc
FDBinary}) assume binarity to be the only cause of the line profile
variations (LPV) detected in the spectra. This implies that
individual stellar components of a binary are supposed to be
intrinsically non-variable, making the application of the method to
our original data set impossible.

\begin{figure*}
\includegraphics[scale=0.95]{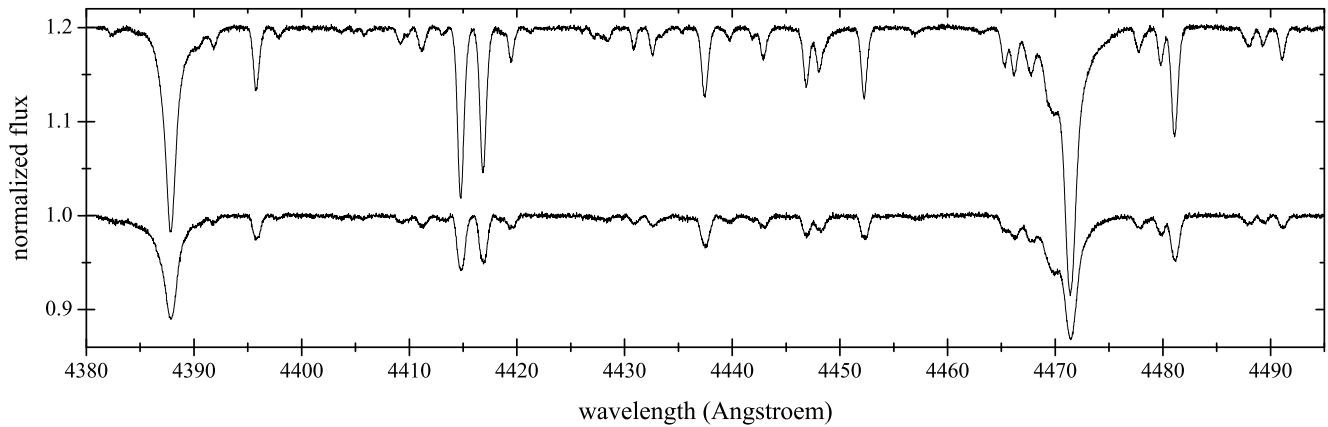}
\caption{A portion of the raw, non-corrected for the individual
light contributions disentangled spectra of both components of the
$\sigma$~Sco binary. The spectrum of the primary was vertically
shifted by a constant value for visibility purposes.}
\label{Figure4}
\end{figure*}

Our first attempt to solve the problem with the intrinsically
variable primary component of $\sigma$\,Sco was to bin the original
spectra with the orbital period determined in Sect.~3. From a set of
experiments, we have found 50 orbital phase bins (corresponding to
an orbital phase bin width of 0.02) to be the optimal choice: this
minimised the number of orbital phase gaps, on the one hand, and
provided sufficient pulsation phase coverage in each bin, on the
other hand. A selection of a larger/smaller bin width is associated
with additional uncertainties due to the displacement of the lines
of the secondary within a given bin, and due to an increased number
of orbital phase gaps, respectively. If the spectra were uniformly
distributed with the (dominant) pulsation period of the primary in
each orbital phase bin, one would expect the effect of the stellar
oscillations to cancel out, providing us with a data set well
suitable for orbit determination and spectral disentangling. In
reality, the spectra do not show such a uniform distribution with
the pulsation phase, and the {\sc spd} technique delivers unreliable
results. We attempted to solve the problem by selecting only those
spectra that deviated from the mean spectrum of the corresponding
bin in RV by less than 2~\kms. At this point we still assume that
the mean spectrum in each bin represents a zero pulsation phase
(unperturbed profile). Although this is not the case, our tight
selection criterion of 2~\kms\ for the individual spectra led to the
remarkable suppression of the amplitude of the dominant pulsation
mode, compared to the binning procedure alone. This selection
resulted in 131 useful spectra out of 1020 that we had in total. The
value of 2~\kms\ is not a random choice but was selected based on a
set of experiments. In particular, we aimed to reduce the influence
of the dominant radial mode on the orbital solution as much as
possible, and, on the other hand, to have enough spectra to provide
sufficient orbital phase coverage, which would make the
determination of precise orbital elements possible. For the
calculation of the orbital elements, we used five spectral intervals
centered at the C\,{\small II}\,4267\,\AA, He\,{\small
I}\,4471\,\AA, He\,{\small I}\,4713\,\AA, He\,{\small I}\,5015\,\AA,
and He\,{\small I}\,5047\,\AA\ lines. The orbital solution obtained
based on the selection of spectra described above is indicated as
``Solution~1'' in Table~\ref{Table5}; the corresponding orbital
elements are listed in the third column. We fixed the period to the
value obtained in Sect.~3, all other parameters were allowed to be
free. The derived values of eccentricity and time of the periastron
passage are consistent with those obtained from the iterative
approach (cf. Table~\ref{Table3}), whereas the longitude of the
periastron and, particularly, the semi-amplitude $K_1$ differ
significantly. After a careful inspection of the disentangled
spectra in the five spectral regions mentioned above, it became
clear that the solution still suffers from the large amplitude
dominant pulsation mode of the primary component. In particular,
some weak ``absorption'' features showed up in the outer wings of
the photospheric stellar lines. The positions and regularity of
these features suggest that they are artificial and have nothing to
do with real absorption in the photosphere of either of the binary
components. To solve the problem, we reduced our set of 131 spectra
even further, by tightening the selection criterion from 2 to
1~\kms. This selection gave us 30 spectra that were finally used for
the spectral disentangling. The bottleneck of such selection is that
the spectra do not provide good enough orbital phase coverage, in
the sense that the regions around the minimum and maximum RV are
badly covered. This makes the determination of the eccentricity
impossible, thus we decided to fix it to the value obtained from the
previous solution that was obtained based on the selection of 131
spectra. The remaining four orbital elements, semi-amplitudes $K_1$
and $K_2$, longitude of the periastron $\omega$ and the time of the
periastron passage $T_0$, were set as free parameters. The orbital
solution obtained this way is indicated as ``Solution~1'' in
Table~\ref{Table5}; the corresponding orbital elements are given in
the fourth column.

\begin{table}
\tabcolsep 1.5mm\caption{Atmospheric parameters and individual
abundances of both components of the $\sigma$~Sco system. Abundances
are expressed relative to $\log N(\rm H) = 12.0$. Cosmic Abundance
Standard (CAS) is taken from \citet{Nieva2012}. {Error bars are
expressed in terms of a 3$\sigma$ level.}}\label{Table6}
\begin{tabular}{llr@{$\pm$}lr@{$\pm$}lr@{$\pm$}l} \hline
Parameter & \multirow{2}{*}{Unit} &
\multicolumn{2}{c}{\multirow{2}{*}{Primary}} &
\multicolumn{2}{c}{\multirow{2}{*}{Secondary}} &
\multicolumn{2}{c}{\multirow{2}{*}{CAS}}\\
\multicolumn{1}{c}{Elem.} & & \multicolumn{2}{}{} & \multicolumn{2}{}{} & \multicolumn{2}{}{}\\
\hline \multicolumn{8}{c}{{\bf
Atmospheric parameters}}\\
\te\rule{0pt}{9pt} & K & 25\,200&1\,500 & 25\,000&2\,400 & \multicolumn{2}{}{}\\
\logg & dex & 3.68&0.15 & 4.16&0.15 & \multicolumn{2}{}{}\\
\vsini & \kms & 31.5&4.5 & 43.0&4.5 & \multicolumn{2}{}{}\\
\vmicro & \kms & 14.0&3.0 & 4.0&3.0 & \multicolumn{2}{}{}\\
\multicolumn{8}{c}{{\bf Individual abundances\rule{0pt}{13pt}}}\\
He & \multirow{7}{*}{dex} & 10.94&0.24 &\multicolumn{2}{c}{$>$10.70} & 10.99&0.01\\
C & & 8.20&0.12 &8.11&0.12 & 8.33&0.04\\
N & & 7.68&0.09 & 7.62&0.21 & 7.79&0.04\\
O & & 8.76&0.21 & 8.79&0.24 & 8.76&0.05\\
Si & & 7.43&0.24 & 7.43&0.36 & 7.50&0.05\\
Mg & & 7.40&0.30 & 7.35&0.36 & 7.56&0.05\\
Al & & 6.07&0.15 & 6.03&0.15 & \multicolumn{2}{c}{}\\ \hline
\end{tabular}
\end{table}

\begin{figure*}
\includegraphics[scale=0.42]{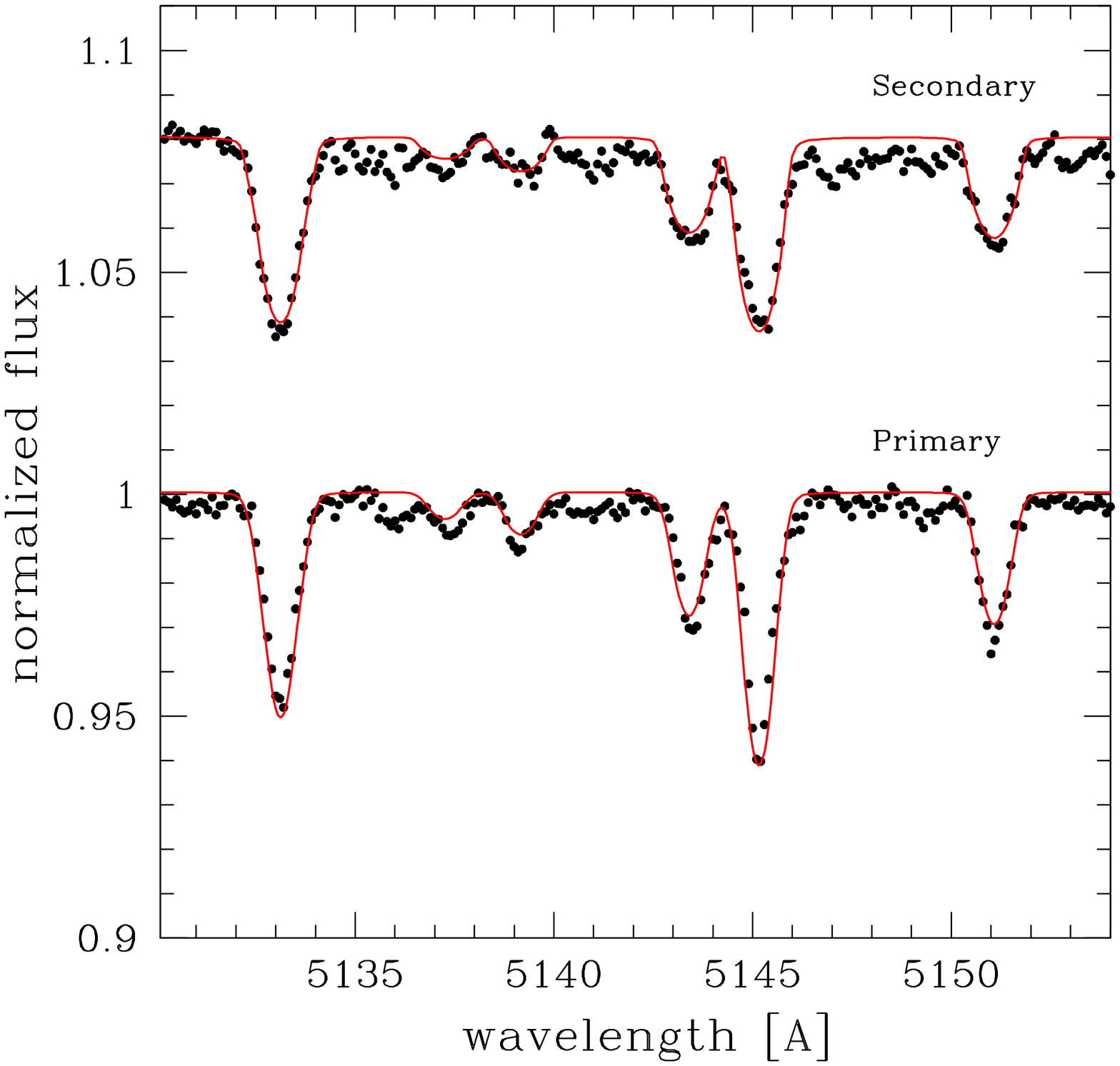}\hspace{10mm}
\includegraphics[scale=0.42]{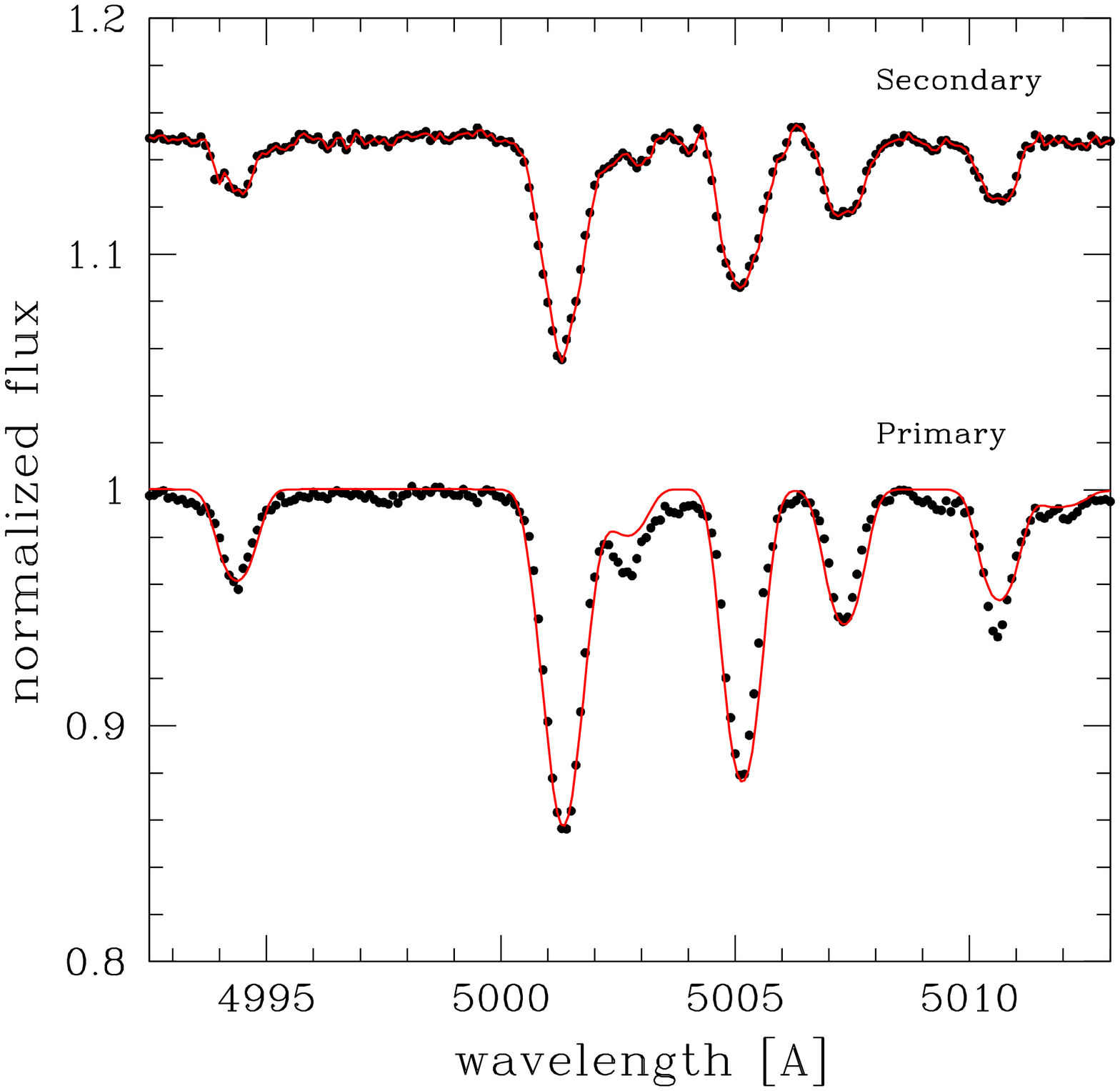}
\caption{Best fit (lines) to a set of carbon (left) and nitrogen
(right) lines in the spectra of both components of the $\sigma$~Sco
binary system.} \label{Figure5}
\end{figure*}

The disentangled spectra obtained from our Solution~1 were used for
a detailed spectrum analysis by means of the methods outlined in
Sect.~4.2. The analysis showed that the obtained orbital solution
assumes a too small contribution of the secondary component to the
total light of the system. In the result, the disentangled spectrum
of the secondary shows an unreliably high abundance of oxygen: we
detected an overabundace above 1~dex compared to the standard cosmic
abundances reported in \citet{Nieva2012}. Such large oxygen
enrichment is not predicted theoretically and has not been
previously detected in any high-mass stars. Moreover,
\citet{Morel2009} shows that OB stars tend to show slightly sub- or
at most solar oxygen abundances. This led us thinking that the
obtained orbital solution is not reliable and this, in particular,
concerns the K$_1$ semi-amplitude of the primary component. Indeed,
the value of $\sim$25~\kms\ that we obtained is very close to the RV
semi-amplitude of the primary's dominant pulsation mode (cf.
Table~\ref{Table4}), and is significantly lower than the value
derived in Sect.~3.2 and reported in \citet{Mathias1991} and
\citet{Pigulski1992}, for example.

In the next step, we fixed the orbital period and RV semi-amplitude
of the primary to the values derived in Sect.~3.2, and re-computed
our orbital solution and the disentangled spectra. This was also
done in two steps: i) using the set of 131 spectra and allowing
eccentricity to be one of the free parameters; and ii) fixing
additionally the eccentricity and fine-tuning K$_2$, $\omega$, and
$T_0$ based on the reduced set of the 30 observed spectra. Both
steps are depicted in Table~\ref{Table5} as ``Solution 2'', where
the fourth column lists the orbital elements that we consider as the
final ones. Except for the value of $\omega$, this orbital solution
agrees within the error bars with our preliminary solution from
Table~3. These orbital elements were used to compute the
disentangled spectra of both stars; a small part of the raw spectra
non-corrected for the individual light contributions are illustrated
in Fig.~\ref{Figure4}. The spectrum of the primary was vertically
shifted for better visibility. From visual inspection of the
spectra, it is obvious that both stars are of similar spectral type
but have (slightly) different rotational velocities. The detailed
spectrum analysis of both stars is outlined in the next section and
includes the estimation of their fundamental parameters and
atmospheric chemical composition.

\subsection{Spectrum analysis of both binary components}

In the case of eclipsing binaries, a combination of high-resolution
spectroscopic data and a spectral disentangling technique allows the
estimation of the effective temperatures and surface gravities as
well as chemical composition of individual binary components to a
very high precision \citep[see,
e.g.,][]{Pavlovski2005,Hareter2008,Pavlovski_Southworth2009,Tkachenko2009,Tkachenko2010}.
Though the accuracy that is reached in the parameters derived from
spectroscopy for the components of non-eclipsing systems is
comparable to the one achieved for single stars, it is still
significantly higher than the accuracy expected from the analysis of
broad-band photometric data, for example.

Before the disentangled spectra can be analysed to determine
properties of stars, they have to be renormalised to the individual
continua according to the light ratio of the stars. This important
parameter can be accurately determined for eclipsing binaries from
photometric data \citep[see,
e.g.,][]{Pavlovski2009,Tkachenko2012,Debosscher2013,Lehmann2013,Hambleton2013}.
When a binary star is not eclipsing or no photometric data is
available in the eclipses, the light ratio is set as a free
parameter and is determined simultaneously with the fundamental
parameters of the components from the disentangled spectra
\citep[the so-called {\it constrained fitting},][]{Tamajo2011}. The
fundamental parameters of both stars were determined by fitting the
disentangled spectra to a grid of synthetic spectra. The grid was
computed based on the so-called {\it hybrid approach}, that is using
LTE-based atmosphere models calculated with the {\sc atlas9} code
\citep{Kurucz1993}, and non-LTE spectral synthesis by means of the
{\sc detail} \citep{Butler1984} and {\sc surface}
\citep{Giddings1981} codes. Justification of such approach is
discussed in \citet{Nieva2007}. The procedure we used to estimate
the fundamental parameters and chemical composition of both
components of the $\sigma$~Sco system is outlined in detail in
\citet{Tamajo2011} and \citet{Tkachenko2014}.

Table~\ref{Table6} lists the atmospheric parameters and individual
abundances of both stellar components of the $\sigma$~Sco system;
the quality of the fit to a set of carbon and nitrogen lines is
shown in Fig.~\ref{Figure5}. The errors on the parameters and
abundances are 3$\sigma$ levels, with $\sigma$ the standard
deviations of the mean. The way we did the selection of the spectra
suitable for the spectral disentangling (see Sect.~4.1 for details)
was such as to use the spectra representative of the equilibrium
stage of the primary component with respect to its dominant radial
pulsation mode. Consequently, the fundamental parameters of the
primary listed in Table~\ref{Table6} are valid for the equilibrium
state of the primary. Note that \citet{Morel2006} investigated the
effect of taking spectra at minimum and maximum radius for the
large-amplitude radial $\beta\,$Cep pulsator $\xi^1\,$CMa and found
a range in \te\ and \logg\ equal to 1000\,K and 0.1 dex,
respectively. The amplitude of the radial mode of $\sigma\,$Sco is
somewhat larger than the one of $\xi^1\,$CMa, so we expect the same
range to apply here. For this reason, we carefully selected spectra
at the equilibrium phase of the primary to deduce the fundamental
parameters. The binary components are found to have similar spectral
types and chemical composition, but differ in luminosity class. For
the majority of the elements, the derived abundances agree within
3$\sigma$ error bars with the Cosmic Abundance Standard (CAS)
determined by \citet{Nieva2012} from the analysis of B-type stars in
OB associations and for field stars in the solar neighbourhood. An
exception is the carbon abundance found in the secondary which is at
$\sim$0.2~dex lower level than the CAS value.

We present a lower limit for the atmospheric helium abundance of the
secondary component only. Several strong He~{\small {\sc I}} lines
that were used to derive the helium content for this star suggested
an abundance of $\sim$10.70~dex, which is significantly lower than
the CAS value. On the other hand, the abundance inferred from the
He~{\small {\sc II}} line at $\lambda\lambda$~4686~\AA\ is
consistent with the CAS value. This discrepancy between the
abundances inferred from He~{\small {\sc I}} and He~{\small {\sc
II}} lines cannot be explained by an erroneous effective temperature
for the secondary, however: the temperature of this star is well
constrained by the visibility of He~{\small {\sc II}} line at
$\lambda\lambda$~4686~\AA\ and the absence of two other He~{\small
{\sc II}} lines at $\lambda\lambda$~4200~\AA\ and
$\lambda\lambda$~4541~\AA\ in its spectrum. Though a temperature
change could account for low helium abundance obtained from the
lines of neutral atoms of this element, it also greatly affects the
visibility of He~{\small {\sc II}} lines and the quality of the fit
of metal and hydrogen lines in the spectrum of this star. Thus, we
conclude that the strong He~{\small {\sc I}} lines present in the
spectrum of the secondary still suffer from the large-amplitude
oscillations of the primary in the composite spectra, and the helium
abundance derived by us for the secondary component can only be
considered as a lower limit.

\begin{table}
\tabcolsep 2.0mm\caption{System and physical parameters of the
$\sigma$~Sco binary.}\label{Table7}
\begin{tabular}{llr@{$\pm$}lr@{$\pm$}l} \hline
Parameter & Unit & \multicolumn{2}{c}{Primary} & \multicolumn{2}{c}{Secondary}\\
\hline orbital inclination$^*$ ($i$) & degrees &
\multicolumn{4}{c}{21.8$\pm$2.3}\\
semi-major axis ($a$) & AU & \multicolumn{4}{c}{0.583$\pm$0.064}\\
ang. semi-major axis$^*$ ($a''$) & mas & \multicolumn{4}{c}{3.62$\pm$0.06}\\
ang. diameter$^*$ ($\theta$) & mas & 0.67&0.03 & 0.34&0.04\\
\hline\vspace{1.3mm}
Mass ($m$)\rule{0pt}{9pt} & M$_{\odot}$ & 14.7&4.5 & 9.5&2.9\\
\multirow{2}{*}{Radius ($R$)} & \multirow{2}{*}{R$_{\odot}$} &
9.2&1.9 & 4.2&1.0\\\vspace{1.3mm}
             &                              &11.3&1.6$^{**}$ &
             5.8&1.2$^{**}$\\
\multirow{2}{*}{Luminosity (log($L$))} & \multirow{2}{*}{L$_{\odot}$} & 4.49&0.24 & 3.80&0.26\\
                      &                              &4.67&0.17$^{**}$ & 4.07&0.26$^{**}$\\
\hline \multicolumn{6}{l}{$^*$ taken from \citet{North2007}}\\
\multicolumn{6}{l}{$^{**}$ computed from angular diameters}
\end{tabular}
\end{table}

\section{System and physical parameters of $\sigma$~Sco}

The fact that $\sigma$~Sco is not an eclipsing binary withholds us
from measuring the orbital inclination angle and thus the masses of
the individual stellar components. We use the orbital inclination
angle determined by \citet{North2007} from interferometry to
estimate the physical parameters of both components of the
$\sigma$~Sco system. We point out that \citet{North2007} did not
take into account the pulsations of the primary in their analysis.

The sum of the semi-major axes of the component orbits is given by
\begin{equation}
a=a_1+a_2=\frac{P(K_1+K_2)\sqrt{1-e^2}}{2\pi \sin i}, \label{Eq.1}
\end{equation}
with $P$ the orbital period in seconds, $K_{1,2}$ the RV
semi-amplitudes of the components, $e$ and $i$ the orbital
eccentricity and inclination angle, respectively.

Kepler's third law reads
\begin{equation}
P^2=\frac{4\pi^2}{G(m_1+m_2)}a^3,
\end{equation}
or
\begin{equation}
M=m_1+m_2=\frac{a^3}{P^2}, \label{Eq.3}
\end{equation}
with $M$ the total mass of the system measured in M$_{\odot}$, and
$a$ and $P$ measured in astronomical units ($AU$) and years,
respectively. Given that
\begin{equation}
\frac{K_1}{K_2}=\frac{a_1}{a_2}=\frac{m_2}{m_1},
\end{equation}
Eqs.~(\ref{Eq.1}) and (\ref{Eq.3}) lead to estimates of individual
masses of both components. If the mass $m$ and surface gravity $g$
of the star are known, the radius and luminosity can be computed
from
\begin{equation}
R=\sqrt{\frac{Gm}{g}}\ \ \ \ \ \ {\rm and}\ \ \ \ \ \
\frac{L}{L_{\odot}}=\left(\frac{R}{R_{\odot}}\right)^2\left(\frac{T}{T_{\odot}}\right)^4
\label{Eq.5}
\end{equation}
with $G$ the gravitational constant. If the angular semi-major axis
($a''$) and angular diameters ($\theta_{1,2}$) of both stars are
known, the component's radii can be alternatively computed from:
\begin{equation}
D = \frac{1}{\pi_d}=\frac{a}{a''};\ \ \ \ \ \ \ \ \ \ \ \
R_{1,2}=D\tan\left(\frac{\theta_{1,2}}{2}\right), \label{Eq.6}
\end{equation}
with $D$ the distance, and $\pi_d$ the dynamical parallax.

Table~\ref{Table7} lists the system and physical parameters of
$\sigma$~Sco. The errors in the parameters arising from the
uncertainties in the orbital elements were computed by repeating the
calculations after varying the elements within their 1$\sigma$
uncertainty levels. Our masses are lower than those reported by
\citet{North2007}, although both estimates agree within the error
bars, which reveal rather bad precisions of $\sim$30\% for the
masses of both stars. The total mass of the system of
30.3$\pm$9.0~M$_{\odot}$ presented by \citet{North2007} is about
25\% larger than the total mass derived in this study. This is
primarily due to the large differences in the assumed semi-amplitude
of the secondary component and the eccentricity of the system. Given
the large error bars, the mass of the primary component of
14.7$\pm$4.5~M$_{\odot}$ derived by us is in agreement with the
majority of members from the catalogue of Galactic $\beta$~Cep stars
\citep{Stankov2005}, where the mass distribution was found to peak
at 12~M$_{\odot}$.

The radii of the components derived from the masses and gravities,
and from the angular diameters of the stars, differ by about 1.9 and
1.6~R$_{\odot}$ for the primary and secondary, respectively. This
difference is within the quoted error bars. \citet{North2007}
reported the radius of 12.7$\pm$1.8~R$_{\odot}$ for the pulsating
primary component and adopted the radius of 6.4~R$_{\odot}$ for the
secondary from its spectral classification. The difference between
our radius of the primary determined from its angular diameter and
the \citet{North2007} value is due to the difference in the
semi-major axis of the system (see above). In Sect.~7, we perform a
detailed asteroseismic modelling of the primary component of
$\sigma$~Sco to verify the determined masses and radii of both
binary components.

\section{Spectroscopic mode identification}

In this section, we describe the results of the spectroscopic mode
identification for the primary of $\sigma$~Sco. We make use of the
orbital solution and the disentangled spectrum of the secondary
obtained in Sect.~4.1 to subtract the contribution of the secondary
component from the observed, composite spectra and to shift the
residual profiles to the reference frame of the primary.

For the extraction of the individual frequencies from both RVs and
line profiles, we used the discrete Fourier-transform (DFT) and the
consecutive prewhitening procedure as implemented in the {\sc
famias} package \citep{Zima2008}. The DFT was computed up to the
Nyquist frequency of the data set, and similar to the results
presented in Sect.~3.2, no significant contribution was found in the
high-frequency domain. At each step of the prewhitening, we
optimized the amplitudes and phases of the individual peaks whereas
the frequency values were kept fixed. We restricted our analysis to
the Si\,{\small III}\,4552~\AA\ spectral line which is known to be
sensitive to stellar oscillations in hot stars \citep{Aerts2003}.

Both the Fourier-parameter fit \citep[FPF,][]{Zima2006} and the
moment \citep{Aerts1992} methods give results consistent with those
presented in Sect.~3.2. Three independent modes with frequencies
f${_1}=4.0513$, f${_3}=4.1727$, and f${_5}=5.9702$~\cd\ were
unambiguously detected with both methods, while the frequency
$f{_6}=0.4997$~\cd\ could be detected in the RVs only and with low
significance level (S/N=3.8). In addition, the orbital frequency and
the second and the third harmonics of the dominant radial mode have
been detected with both diagnostics.

In the next step, we used all our 1020 spectroscopic measurements to
identify two out of three detected frequencies in terms of $l$ and
$m$ quantum numbers. We fixed $l=0$ for the f${_1}=4.0513$~\cd\
mode, according to previous studies
\citep[e.g.,][]{Cugier1992,Heynderickx1994} and characteristic for
radial modes displacements of the oscillation sensitive spectral
lines in the course of the pulsation cycle. This allows us to limit
the parameter range to two independent modes during mode
identification and to speed up the calculations.

Similar to the frequency analysis, the mode identification is
performed using two different approaches, the FPF and the moment
methods. Both rely on the calculation of the synthetic profiles but
use different observables to deliver the identification of the
individual pulsation modes. In the FPF method, the observed Fourier
parameters zero point, amplitude, and phase across the line profile
are compared to the theoretical values, while in the moment method,
one relies on the integrals of the line profile \citep{Zima2008}. To
provide sufficient spatial resolution, we divide the stellar surface
into 10\,000 segments. The required fundamental parameters like
mass, radius, etc. were set according to our best fitting model
presented in Table~\ref{Table8} (see also Sect.~7 for details).

The moment method delivers unambiguous mode identifications for
f${_3}$ and f${_5}$ as being ($l,m$)=(1,1) and a radial modes,
respectively. Our ten best solutions provide the same identification
for these two modes but disagree on the inclination angle of the
rotation axis to the line-of-sight and the amplitude of the f$_{5}$
mode. The inclination angle varies between 17 and 32 degrees, and we
cannot distinguish between different solutions due to rather small
changes in the $\chi^2$ value. Given the range in the inclination
angle, the primary of $\sigma$~Sco has equatorial rotation velocity
between 60 and 110~\kms. Using the mass and radius derived in the
next section, we find the primary to rotate between 13.5 and 25~\%
of its break-up velocity. The orbital inclination angle of
21.8$\pm$2.3 degrees reported by \citet{North2007} is within the
range we derived for the inclination angle of the rotation axis of
the primary. This suggests spin-orbit alignment or at most slight
misalignment for the primary component of the $\sigma$~Sco system.

The results of the mode identification obtained with the FPF method
are less satisfactory, however. A variety of solutions with nearly
the same goodness of the fit was obtained. For both fitted modes,
f${_3}$ and f${_5}$, the spherical degree $l$ was found to range
between 0 and 3, with a slight preference for f${_3}$ being $l=1$ or
2 mode. It is well known that the moment method is better suitable
for slow to moderate rotators, while sufficiently large rotational
broadening of the lines is required for the FPF method to work
properly. The limitations we encountered in using the FPF technique
for the mode identification is probably due to the insufficient
rotational broadening of the lines of the primary.

\section{Asteroseismic modelling}

Asteroseismic modelling of massive stars is done by adapting a
forward modelling approach for frequency matching, starting from a
set of models in the spectroscopic error box (\te, \logg) of the
star \citep[][Chapter~7]{Aerts2010}. We use the {\sc mesa} stellar
structure and evolution code developed by
\citet{Paxton2011,Paxton2013} to compute non-rotating models. This
is justified in view of the relatively modest rotation of less than
a quarter of the critical velocity. The initial abundance fractions
(X, Y, Z) = (0.710, 0.276, 0.014) are those from \citet{Nieva2012},
in agreement with the spectroscopic findings (cf. Sect.~4).
Convective core overshoot is described in exponentially decaying
prescription of \cite{Herwig}. The Ledoux criterion is used in the
convection treatment. The OPAL opacity tables \cite{Opal} and MESA
equation-of-state are used.

\begin{figure*}
\includegraphics[scale=3.5]{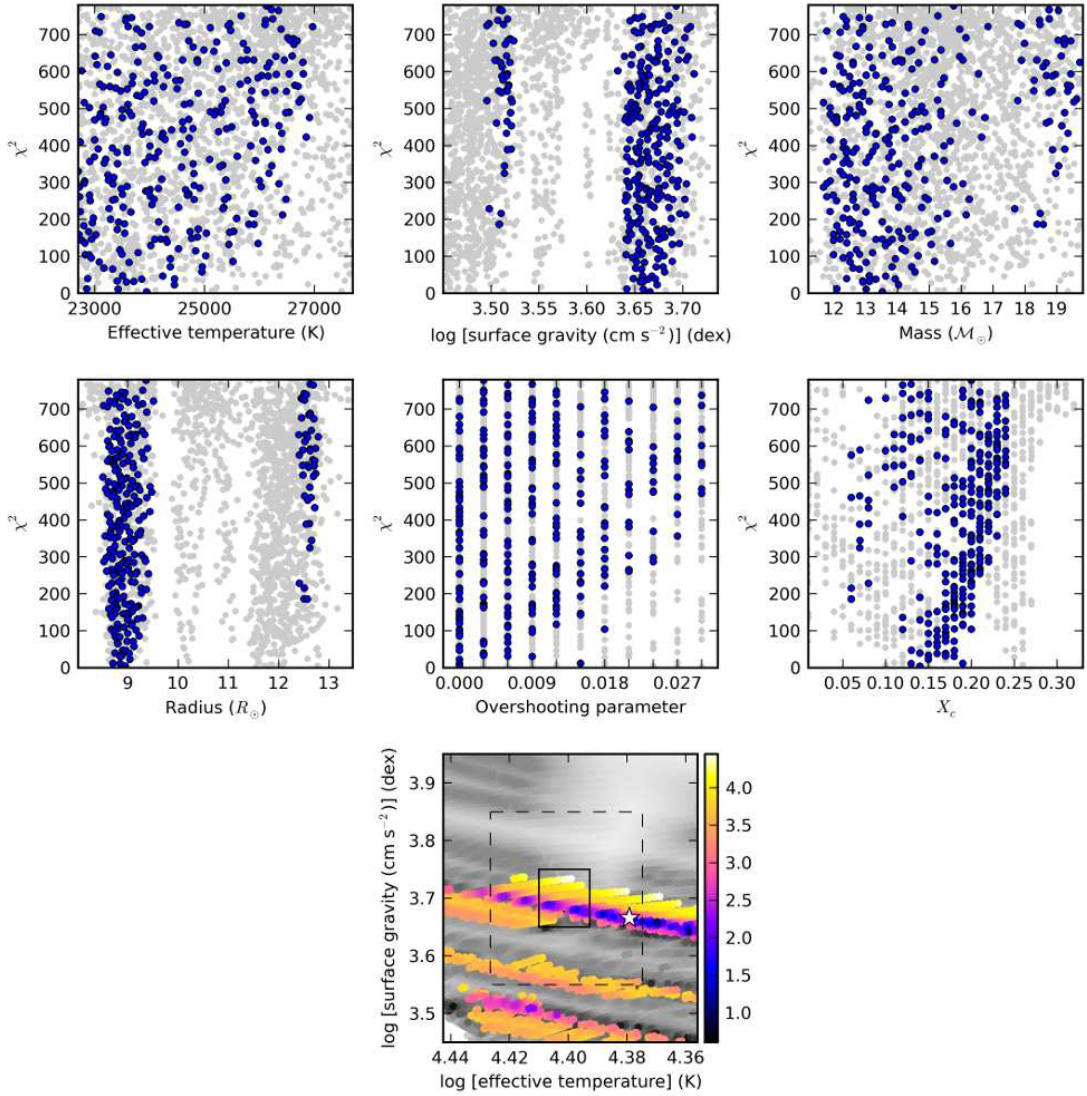}
\caption{{\bf Top, Middle:} $\chi^2$-distributions for six stellar
model parameters. The background light gray and foreground blue dots
in each panel show distributions obtained without putting
constraints on the identification of the dominant pulsation mode and
assuming that it is a radial mode, respectively. {\bf Bottom:}
Position of the best fit theoretical model (asterisk) and 1- and
3-sigma observational error boxes (solid and dashed lines,
respectively) in the $\log$\te--\logg\ plane. The $\chi^2$ value is
colour coded. See text for more details.} \label{Figure6}
\end{figure*}

We screened the mass range from 10 up to 20~M$_{\odot}$ to interpret
the pulsations of the primary component of $\sigma$~Sco. The central
hydrogen content (X$_c$) and the overshooting parameter ($f_{ov}$)
were also varied in the ranges from 0.7--0.0 and 0.0--0.03,
respectively, which corresponds with our step-wise overshoot
parameter of $\alpha_{ov}$~$\epsilon$ [0.0,0.3]~H$_{\rm p}$. All
models fitting the observed spectroscopic values of \te\ and \logg\
within 5-sigma have been selected. This selection resulted in some
28\,000 evolutionary models whose p- and g-mode eigenfrequencies for
mode degrees $l=0$ to 3 have been calculated in the adiabatic
approximation with the the {\sc gyre} stellar oscillation code
\citep{Townsend2013}. The theoretical frequencies were compared with
the three observed dominant modes of the primary of $\sigma$~Sco
(f$_1$, f$_3$, and f$_5$; cf. Table~\ref{Table4}). The best fitting
model was selected using a $\chi^2$ criterion. Figure~\ref{Figure6}
(light gray dots in the top and middle panels) illustrates
$\chi^2$-distributions for six stellar parameters (from top left to
middle right: effective temperature, surface gravity, mass, radius,
overshooting parameter, and central hydrogen content) for the above
mentioned $\sim$28\,000 models. The effective temperature, mass, and
overshooting parameter are poorly constrained, and more than one
minimum is found in the distributions of the surface gravity,
radius, and central hydrogen content.

\begin{figure*}
\includegraphics[scale=1.0]{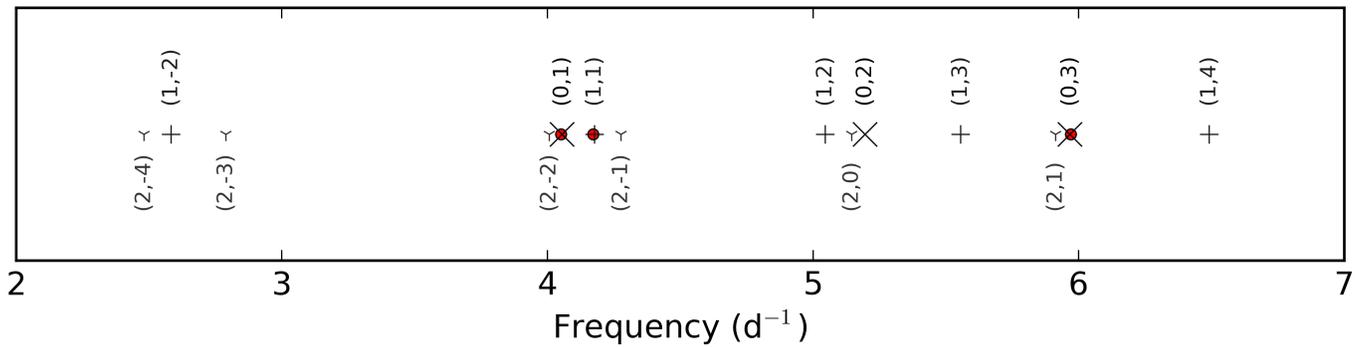}
\caption{Comparison between the observed (red circles) and
theoretical frequencies. Crosses, pluses, and triangular markers
refer to $l$ = 0, 1, and 2 modes, respectively. Each symbol is
assigned to two numbers which correspond with the spherical degree
$l$ and radial order $n$.} \label{Figure7}
\end{figure*}

\begin{table*}
\tabcolsep 2.0mm\caption{Parameters of the three best fitting
stellar evolution models. See text for details.}\label{Table8}
\begin{tabular}{cr@{ }lccccccr@{.}l} \hline
Model & \multicolumn{2}{c}{Temperature (\te)} & Gravity (\logg) & Mass (M) & Radius (R) & overshoot ($f_{ov}$) & Hydrogen content (X$_c$) & Age & \multicolumn{2}{c}{Reduced}\\
 & \multicolumn{2}{c}{K} & \multicolumn{1}{c}{dex} & \multicolumn{1}{c}{M$_{\odot}$} & \multicolumn{1}{c}{R$_{\odot}$} & \multicolumn{1}{c}{H$_{\rm p}$} & \multicolumn{1}{c}{mass fraction} & Myr & \multicolumn{2}{c}{$\chi^2$}\\\hline
1 & 23&945$^{+500}_{-990}$ & 3.67$^{+0.01}_{-0.03}$ & 13.5$^{+0.5}_{-1.4}$ & 8.95$^{+0.43}_{-0.66}$ & 0.000$^{+0.015}$ & 0.14$^{+0.05}_{-0.01}$ & 12.1$^{+2.0}_{-1.0}$ & 4&033\\
2\rule{0pt}{11pt} & 23&430$^{+1020}_{-580}$ & 3.66$^{+0.02}_{-0.02}$ & 13.0$^{+1.0}_{-0.9}$ & 8.85$^{+0.53}_{-0.56}$ & 0.000$^{+0.015}$ & 0.13$^{+0.06}$ & 12.9$^{+1.2}_{-1.8}$ & 10&980\\
3\rule{0pt}{11pt} & 22&855$^{+1600}$ & 3.64$^{+0.04}$ &
12.1$^{+1.9}$ & 8.70$^{+0.35}$ & 0.015$_{-0.015}$ & 0.19$_{-0.06}$ &
13.6$^{+0.5}_{-2.5}$ & 11&496\\\hline
\end{tabular}
\end{table*}

In the next step, we made use of the fact that the dominant mode
f$_1$=4.0515~\cd\ (46.8759~\mhz) of the primary component is a
radial mode: in our frequency fitting process, we insisted that the
dominant mode has a spherical degree $l=0$. We do not make use of
the identification for f$_3$ and f$_5$, but rather use the results
of the previous section as an a posteriori check. The additional
seismic constraint for f$_1$ allowed to reduce the total number of
models down to $\sim$1700; the corresponding $\chi^2$-distributions
are presented in Figure~\ref{Figure6} (top and middle panels) by the
foreground blue dots. Restriction to the dominant mode being radial
allows us to better constrain all six parameters: clear minima are
defined in \logg, mass, radius, and X$_c$; upper limits can be set
for \te\ ($\sim$24\,500~K) and overshooting parameter $f_{ov}$
($\sim$0.015 pressure scale height). Table~\ref{Table8} lists the
parameters and $\chi^2$ values of three best fitting models;
uncertainties were computed by taking into account 12 models with
$\chi^2$ values below 50. Our lowest $\chi^2$ model suggests a mass
of 13.5~M$_{\odot}$ and a radius of 8.95~R$_{\odot}$ for the
primary, well within the 1-sigma errors from the values derived from
a combination of our spectroscopic parameters and interferometric
orbital inclination (cf. Table~\ref{Table7}). The effective
temperature agrees with the spectroscopic value within 3-sigma
(Fig.~\ref{Figure7}, bottom panel), the surface gravity is in
excellent agreement. The radius of the primary of 11.3~R$_{\odot}$
determined from the angular diameter presented by \citet{North2007}
is too large, however.

Figure~\ref{Figure7} illustrates quality of the fit between the
observed and theoretical frequencies based on our best fit
evolutionary model. The observed values are shown by red dots,
theoretical frequencies are presented by crosses ($l=0$), pluses
($l=1$), and triangular markers ($l=2$). The dominant mode
f$_1$=4.0515~\cd\ (46.8759~\mhz) of the primary is identified as the
fundamental radial mode, frequencies f$_3$=4.1726~\cd\
(48.2770~\mhz) and f$_5$=5.9706~\cd\ (69.0201~\mhz) are found to be
the $l=1$ non-radial mode and the second overtone radial mode,
respectively, which is fully consistent with the spectroscopic mode
identification of the two low-amplitude modes.

\section{Discussion and Conclusions}

In this paper, we presented a detailed spectroscopic study of the
close pair of the $\sigma$~Sco quadruple system. The analysis was
based on some 1\,000 high-resolution, high S/N spectra taken during
three consecutive years, in 2006, 2007, and 2008.

We used two different approaches to determine the precise orbital
elements of the close pair of two B-type stars. The first method
relies on the iterative approach comprising of the determination of
the orbital elements, the analysis of the residuals to extract the
pulsation frequencies, and the cleaning the original data from the
pulsation signal. The cycle was repeated until we could get to a
self-consistent set of orbital elements and pulsation frequencies of
the primary. The second approach was based on the spectral
disentangling technique, where the step of RV determination is
overcome and orbital elements are optimized together with the
disentangled spectra of both binary components. We found that this
latter step suffers from the large-amplitude RV variations intrinsic
to the primary component and it was impossible to obtain reliable
disentangled spectra without a priori assumption on the RV
semi-amplitude $K_1$ of the primary. This assumption was made from
the iterative approach outlined above; the obtained disentangled
spectra were analysed to determine fundamental parameters and
atmospheric chemical composition of both binary components.

The spectrum analysis was based on the non-LTE spectral synthesis
and revealed two stars of the same effective temperature (spectral
type) but different surface gravity (luminosity class). The primary
was found to contribute about 62\% to the total light of the system;
the individual components have a similar atmospheric chemical
composition consistent with the present-day cosmic abundance
standard \citep{Nieva2012}. The spectroscopic flux ratio of
$\sim$1.63 agrees within 1- and 2$\sigma$ error bars with the
luminosity ratio from Tables~\ref{Table7} and \ref{Table9},
respectively. We found a low helium abundance for the secondary
component which we do not trust to be real and attribute it to the
problems we encountered during the disentangling of the spectrum of
the secondary in the regions of strong He~{\small {\sc I}} lines.
This is the first time that detailed abundance analysis is presented
for both components of the $\sigma$~Sco binary, and the first time
that precise fundamental parameters are determined for the secondary
component. The effective temperature of the secondary is in
agreement with the value inferred by \citet{North2007} from the
spectral type of the star. For the primary, our parameters are in
agreement with most recent spectroscopic estimates
\citep[e.g.,][]{Niemczura2005}.

\begin{table}
\tabcolsep 2.0mm\caption{Fundamental parameters of both components
of the $\sigma$~Sco system, after seismic modelling.}\label{Table9}
\begin{tabular}{llll} \hline
Parameter & Unit & \multicolumn{1}{c}{Primary} & \multicolumn{1}{c}{Secondary}\\
\hline
Mass ($m$) & M$_{\odot}$ & 13.5$^{+0.5}_{-1.4}$ & 8.7$^{+0.6}_{-1.2}$ \\
Radius ($R$)\rule{0pt}{11pt} & R$_{\odot}$ & 8.95$^{+0.43}_{-0.66}$ & 3.90$^{+0.58}_{-0.36}$\\
Luminosity (log($L$))\rule{0pt}{11pt} & L$_{\odot}$ & 4.38$^{+0.07}_{-0.15}$ & 3.73$^{+0.13}_{-0.15}$\\
Age of the system\rule{0pt}{11pt} & Myr &
\multicolumn{2}{c}{12.1$^{+2.0}_{-1.0}$}\\\hline
\end{tabular}
\end{table}

Our spectroscopic data revealed three independent pulsation modes
intrinsic to the primary component. The two dominant modes
f$_1$=4.0515~\cd\ (46.8759~\mhz) and f$_3$=4.1726~\cd\
(48.2770~\mhz) were previously reported in several studies and
identified as a radial \citep[e.g.,][]{Cugier1992,Heynderickx1994}
an $l$ = 2 non-radial modes \citep{Cugier1992}, respectively. The
lowest amplitude mode f$_5$=5.9706~\cd\ (69.0201~\mhz) detected in
our spectroscopic data is in excellent agreement with the one found
by \citet{Jerzykiewicz1984} in the differential $uvby$ photometry.
All other variability detected in the spectral lines of the primary
component occurs either at harmonics of the dominant pulsation mode
or at low-significance frequencies, which we do not consider as real
ones.

By subtracting the contribution of the secondary from the composite
spectra and by correcting the residuals for the orbital motion of
the primary, we were able to perform spectroscopic mode
identification for the two modes f$_3$ and f$_5$ (with f$_1$ being
postulated to be a radial mode) of the primary of $\sigma$~Sco. The
identification was done with two different methods,
Fourier-parameter fit and moment method. The former technique seems
to be inapplicable to our data, probably due to the insufficient
rotational broadening of the spectral lines. The moment method
delivers unambiguous identification for both modes, delivering
($l,m$)=(1,1) for f$_3$ and a radial mode for f$_5$. The inclination
angle of the rotation axis of the primary with respect to the
line-of-sight ranges from 17 to 32 degrees. Combined with the
orbital inclination reported by \citet{North2007}, this suggests
alignment or at most a small misalignment for the primary component
of $\sigma$~Sco. With the measurement of the inclination angle of
the primary, we are able to constrain its equatorial rotation
velocity to be between 60 and 110~\kms, implying that the star
rotates between 13.5 and 25\% of its break-up velocity.

\begin{figure}
\includegraphics[scale=1.]{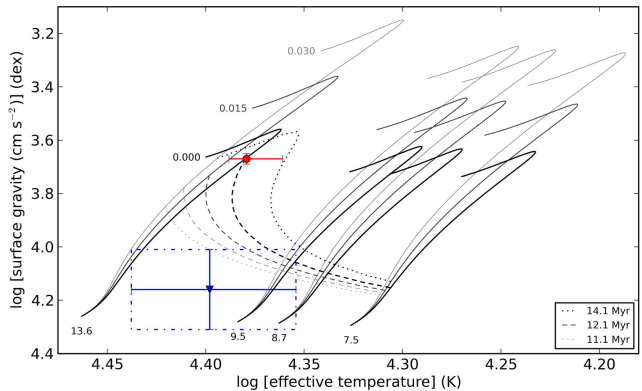}
\caption{Position of the primary (circle) and the secondary
(triangle) of $\sigma$~Sco system in the \te-\logg\ diagram, along
with the {\sc mesa} evolutionary tracks. The initial masses as well
as the overshoot parameter $f_{ov}$ are indicated in the plot. The
atmospheric parameters \te\ and \logg\ of the primary and secondary
are those from asteroseismic and spectroscopic analysis,
respectively. The isochrones corresponding to the age of the system
of 12.1~Myr deduced from seismology of the primary are indicated by
the dashed lines. The error range in age is given by the dotted
lines, which correspond to the best fit overshoot parameter of the
primary $f_{ov}=0.0$.} \label{Figure8}
\end{figure}

We used our orbital elements together with the (interferometric)
orbital inclination angle estimate from \citet{North2007} to compute
individual masses and radii of both components of $\sigma$~Sco. The
values of the radii obtained this way agree within the estimated
errors with those computed from angular diameters presented by
\citet{North2007}. In order to verify the mass and radius of the
primary component, we performed the fitting of all three independent
pulsation modes, using {\sc mesa} and {\sc gyre} -- stellar
evolution and oscillation codes, respectively. Our best fitting
theoretical model delivers an effective temperature and surface
gravity in good agreement with the spectroscopic values, and
confirms the radius of the primary determined from our orbital
solution. The age of the system is estimated to be $\sim$12~Myr.
Table~\ref{Table9} summarizes the final fundamental parameters of
both components of $\sigma$~Sco, where the mass and radius of the
secondary were computed from the theoretical mass of the primary and
spectroscopic mass ratio, and from the calculated mass and
spectroscopic \logg, respectively. The 1$\sigma$ errors given in
Tables~\ref{Table8} and \ref{Table9} do not take into consideration
systematic uncertainties connected with the choice of a given
stellar evolution code and its input physics. Fig.~\ref{Figure8}
shows the position of both components of the $\sigma$~Sco system in
the \te-\logg\ diagram, along with the evolutionary tracks. The
error bars are those obtained from the evolutionary models and
3$\sigma$ spectroscopic uncertainties for the primary and secondary,
respectively. The primary is an evolved star near the end of its
main-sequence, whereas the secondary just started its main-sequence
evolution. The position of the secondary in the diagram suggests
that the star is more massive than 8.7~M$_{\odot}$, the mass we
obtained from binary dynamics and the theoretical mass of the
primary. Similar discrepancy was found for the main-sequence
secondary component of the V380\,Cyg binary system
\citep{Tkachenko2014}.

Our asteroseismic analysis of the primary of $\sigma$~Sco delivered
the identification of all three fitted frequencies. This way, two of
them, f$_1$ and f$_5$, are found to be fundamental and second
overtone radial modes, respectively, and frequency f$_3$ is
identified as an $l=1$ non-radial mode. This is in perfect agreement
with the spectroscopic mode identification we obtained for f$_3$ and
f$_5$ using the moment method. Identification for f$_1$ is
consistent with the majority of the previous studies
\citep[e.g.,][]{Cugier1992,Heynderickx1994} whereas the
identification of f$_3$ is different from the one previously found
by \citet{Cugier1992}. The mode identification for f$_5$ presented
in this study is done for the first time. The addition of the
seismic constraints to the modelling implied a drastic reduction in
the uncertainties of the fundamental parameters, and provided an age
estimate. Our best fitting evolutionary model suggests a small
decrease of 7.3$\times$10$^{-5}$~\cd/century of the frequency of the
dominant radial mode; the effect cannot be detected from our data
due to low frequency resolution.

This paper is the second one in a series devoted to the analysis of
spectroscopic double-lined binary systems, consisting of two B-type
stars of which at least one is a pulsating star. We find a fully
consistent solution and agreement with stellar models for the
primary component, after having taken into account its pulsational
behaviour in the analysis of the data. This conclusions are in sharp
contrast with the incompatibility between data and models for the
binary V380\,Cyg, for which we found a discrepancy of
$\sim$1.5~M$_{\odot}$ between the dynamical and theoretical mass and
a huge core overshoot was needed to explain the properties of the
primary. On the other hand, the fact that V380\,Cyg is an eclipsing
double-lined spectroscopic binary allowed us to measure the masses
of individual component to precisions approaching 1\%, while
precisions in masses of both components of the $\sigma$~Sco system
are as bad as $\sim$30\% if we do not consider the seismic
properties of the primary. In this respect, the two systems are
incomparable and purely measured dynamical masses of the
$\sigma$~Sco stellar components prevent us from drawing any firm
conclusions with respect to the theoretical models. In our next
paper, we will present a detailed analysis of the Spica system,
based on the space-based photometric and ground-based
high-resolution spectroscopic data.

\section*{acknowledgements}
The research leading to these results has received funding from the
European Community's Seventh Framework Programme FP7-SPACE-2011-1,
project number 312844 (SPACEINN), and from the Fund for Scientific
Research of Flanders (FWO), Belgium, under grant agreement
G.0B69.13. Mode identification results with the software package
{\sc famias} developed in the framework of the FP6 European
Coordination Action HELAS (http://www.helas-eu.org/).

{}

\label{lastpage}

\end{document}